\documentclass[12pt]{article}
\usepackage{amsmath}
\usepackage{graphicx,psfrag,epsf}
\usepackage{enumerate}
\usepackage{natbib}
\usepackage{hyperref} 
\usepackage{xcolor}
\usepackage{float}
\usepackage{authblk}
\newcommand{\blind}{0}

\begin{document}

\def\spacingset#1{\renewcommand{\baselinestretch}%
{#1}\small\normalsize} \spacingset{1}


\if0\blind
{
  \title{\bf Algorithm for detection of illegal discounting in North Carolina Education Lottery}
  \author{Jiayi Fu\\
    Jack B  Prothero \\
    Jan Hannig 
    \thanks{\textit{Jan Hannig's research was supported in part by the National Science Foundation under Grant No. DMS-1916115, 2113404, and 2210337.}}\hspace{.2cm}\\
    The University of North Carolina at Chapel Hill\\}
  \maketitle
} \fi

\if1\blind
{
  \bigskip
  \bigskip
  \bigskip
  \begin{center}
    {\LARGE\bf Title}
\end{center}
  \medskip
} \fi

\bigskip
\begin{abstract}
The lottery is a very lucrative industry. Popular fascination often focuses on the largest prizes. However, less attention has been paid to detecting unusual lottery buying behaviors at lower stakes. Our paper introduces a new model to detect illegal discounting in the North Carolina Education Lottery using statistical analysis of net gains and ticket buying habits. Nine outlying players are flagged and are further examined using a proposed stochastic model to calculate the range of their possible losses in the lottery. The unusual buying patterns of the players flagged as outliers are further confirmed using a K-means clustering analysis of lottery store visiting behaviors. 
\end{abstract}

\noindent%
\vfill

\newpage
\spacingset{1.45} 
\section{Introduction}
\label{sec:intro}

The North Carolina Education Lottery (NCEL) is a thriving business, with sales of \$2.86 billion in 2019. We periodically see unimaginably large jackpots covered in the news. It is normal for some lucky players to win a single large prize in the lottery, but it is unlikely for any given player to win multiple large prizes. North Carolina law (N.C.G.S. §18C) dictates that any prizes exceeding \$600 must be redeemed at a state approved facility and certain information about the winner are considered public record, while smaller prizes can be redeemed in person without creating a permanent record. Therefore we will refer to a prize over \$600 as \textit{recorded prize}. Based on published prize probabilities of 42 randomly sampled scratch-off games, the likelihood of winning a recorded prize is 0.0004 on average. In other words, to win a prize of over \$600, a person would on average have to buy 2500 scratch-off tickets. 
If a lottery player owes back taxes, child support, or some other public debt, the winnings would be used first to satisfy this liability (e.g., N.C.G.S. §18C-134). Therefore such a person might illegally choose to sell a winning ticket to another person at a discount in order to avoid the government garnishing the winnings~\cite{off_bell_2016}. In a recent case, both a father and son were found guilty of engaging in lottery ticket discounting, which amounted to over 20 million dollars in illegally claimed lottery tickets~\cite{Kevin_2023}. However, high-volume lottery players that are not involved in such illicit schemes may also win many prizes over \$600. In the short term, individuals may experience luck in winning multiple prizes. However, in the long term, they would incur losses as they regress to the expected lottery return rate. Our goal is to propose a way to help distinguish between discount ticket purchasers and regular high-volume lucky players.
Recent articles have developed total net winnings estimation techniques for people with a large number of recorded prizes  as part of similar efforts to distinguish high-volume lottery players from people with more sinister intentions. In \cite{arratia2015some} the authors find high-probability lower bounds for total net winnings using an optimization approach. They deduce that a particular lottery player in Florida who won 252 large prizes across many games over the course of three years would have had to spend at least \$2 million if all the tickets were purchased fairly. This finding was alarming enough to draw the attention of law enforcement. \cite{stong2020optimal} find similarly high minimum loss lower bounds when playing repeated-draw games like Pick 4 even with optimal betting strategies. However, these papers made no mention of the impact of small lottery prizes. Consistently winning small prizes could give the player the impression that they are not losing as much, as some of the smallest wins might be immediately used to purchase more lottery tickets (practice called ``reinvesting'' among habitual players). Moreover, the majority of lottery-winning prizes consist of small prizes. For example among the 42 randomly sampled scratch-off games the probability of winning less than \$600 is significantly higher than the probability of winning a recorded prize. Therefore, there is a lot of uncertainty due to the effect of small prizes when estimating a player's net loss incurred in order to win a certain number of times. In this paper, we propose a simulation based approach for estimating potential spread of small prize winnings based on the actual revenue distribution in North Carolina lottery games.
Finally,  according to \cite{guryan2005lucky}, "consumers appear to form habits of where they shop." Therefore, if a person is engaging in illegal discounting, that person will be claiming prizes from a much wider range of stores than a single legitimate player. Thus we combine two approaches to identify suspicious players. First, we estimate the total amount one must spend to win many recorded prizes. Second, we identify people with an unusual distribution of stores where they purchased their winning tickets. Hopefully by combining these two approaches we avoid flagging out players who are truly legitimate high-volume players.

\section{Data}

Via a Freedom of Information Act request on \date{March 20, 2020}, we received the data from North Carolina Education Lottery officials. They contain information about winning lottery prizes above \$600 from 597 North Carolina Education lotteries from March 31st, 2006, to January 31st, 2020, just before the U.S. COVID pandemic outbreak. People with a winning prize of less than \$600 can directly declare at the point of purchase without providing identification information. People who win more than \$600 must go to regional claims centers and fill out an NCEL winner claim form.
Our data is built upon the NCEL winner claim form. Therefore, each row of our data is a single recorded prize. For each recorded prize, we have the following features: the winner’s full name, city, county, game type, prize amount, lottery name, declared place, paid date, selling retailer name, and selling retailer address. Though this data provides winners’ names, we anonymize them in this paper. In all, 391,791 winning prizes were recorded, with 197,930 unique winners collecting these prizes.

\subsection{Data visualization}
We present several plots investigating overall patterns among players who have won large prizes. Recall that our dataset only contains information about recorded wins (wins over \$600). Therefore whenever we refer to a win in this section, we specifically mean a recorded win.

\begin{figure}[H]
\centering
\includegraphics[width=0.8\textwidth]{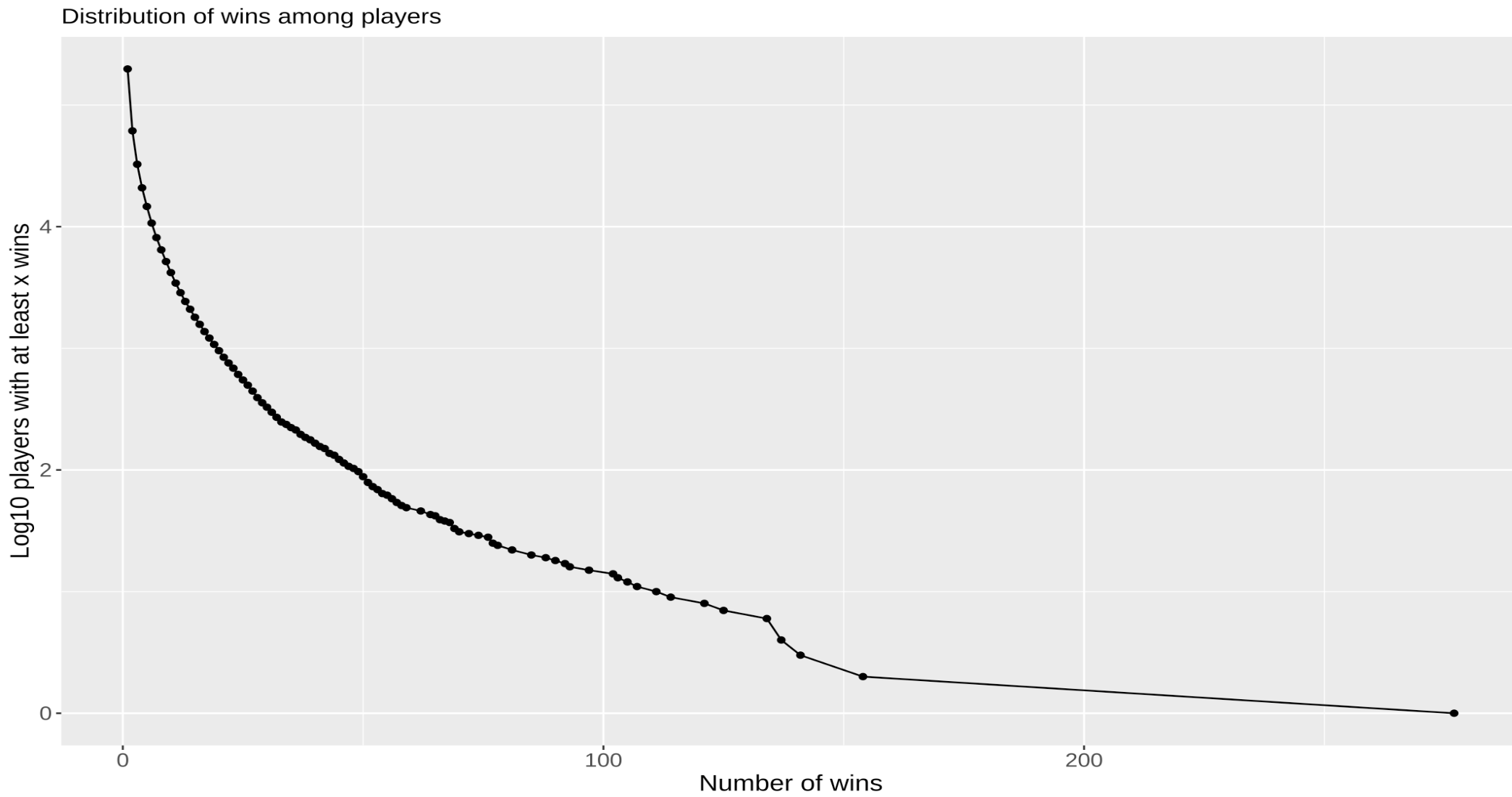}
\caption{{Scatter plot of $\log_{10}$ of people who won at least x times (y axis) vs the number of wins (x axis). Each dot corresponds to the total number of players who won a big prize at least x times.}}
\label{fig:WinDisGraph}
\end{figure}

To get an overall sense of how the 391,791 wins are distributed among the 197,930 players in the data set, we generated a graph that visually represents that distribution. Note that the vertical axis is logarithmic in Figure~\ref{fig:WinDisGraph} as the vast majority of players in the data set have very few wins.
As the number of wins increases, there is a significant decline, with only approximately 1,000 individuals achieving six or more big wins. Fewer than 100 individuals managed to secure at least 49 big wins. The highest number of wins for prizes over \$600 is recorded at 277, marking an exceptional outlier within the dataset. Our primary interest is mostly in players towards the higher end of this distribution and our main task is distinguishing legitimate high-volume players from discounters. 

We also conducted a preliminary investigation into the types of lotteries that have the highest number of significant prizes. The following graph illustrates the lotteries with most number of recorded prizes.
\begin{figure}[H]
\centering
\includegraphics[width=0.8\textwidth]{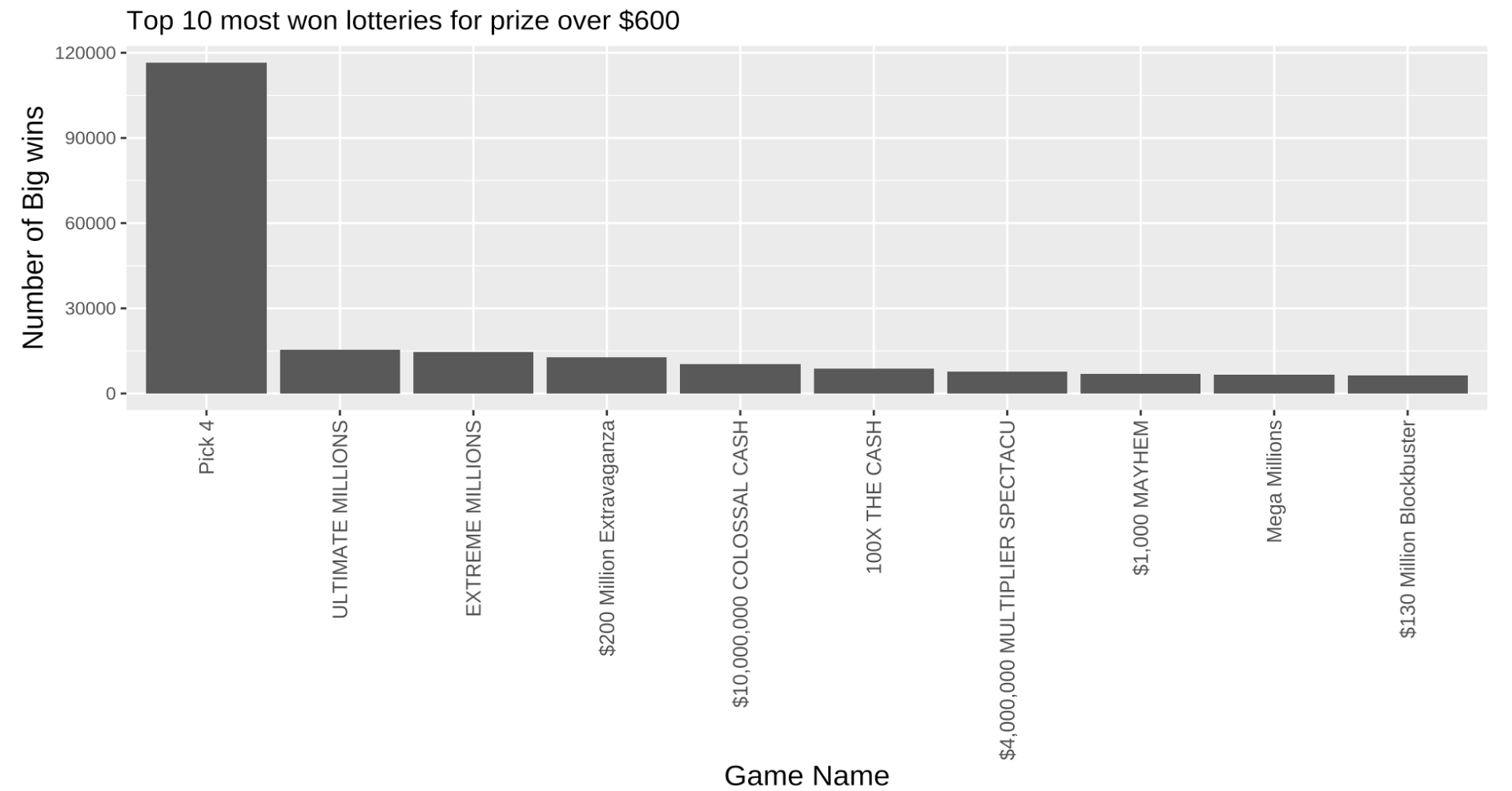}
\caption{Top 10 lottery games by number of recorded wins}
\label{fig:BigWinGameGraph}
\end{figure}
The lottery with the highest number of recorded wins is Pick 4, followed by a mix of online and scratch-off lottery games. The reason for Pick 4's high number of wins is due to the relatively favorable odds of winning a prize over \$600, which are set at 1 in 10,000.

When we started this study, we held an assumption that most lottery players, and especially habitual players, had a small number of favorite stores where they purchased tickets. This assumption is a key component of our proposed method for discriminating between legitimate players and discounters. Therefore, we also explore the distribution of the number of stores at which each player won a big prize in the same manner as the distribution of the number of wins. \begin{figure}[H]
\centering
\includegraphics[width=0.8\textwidth]{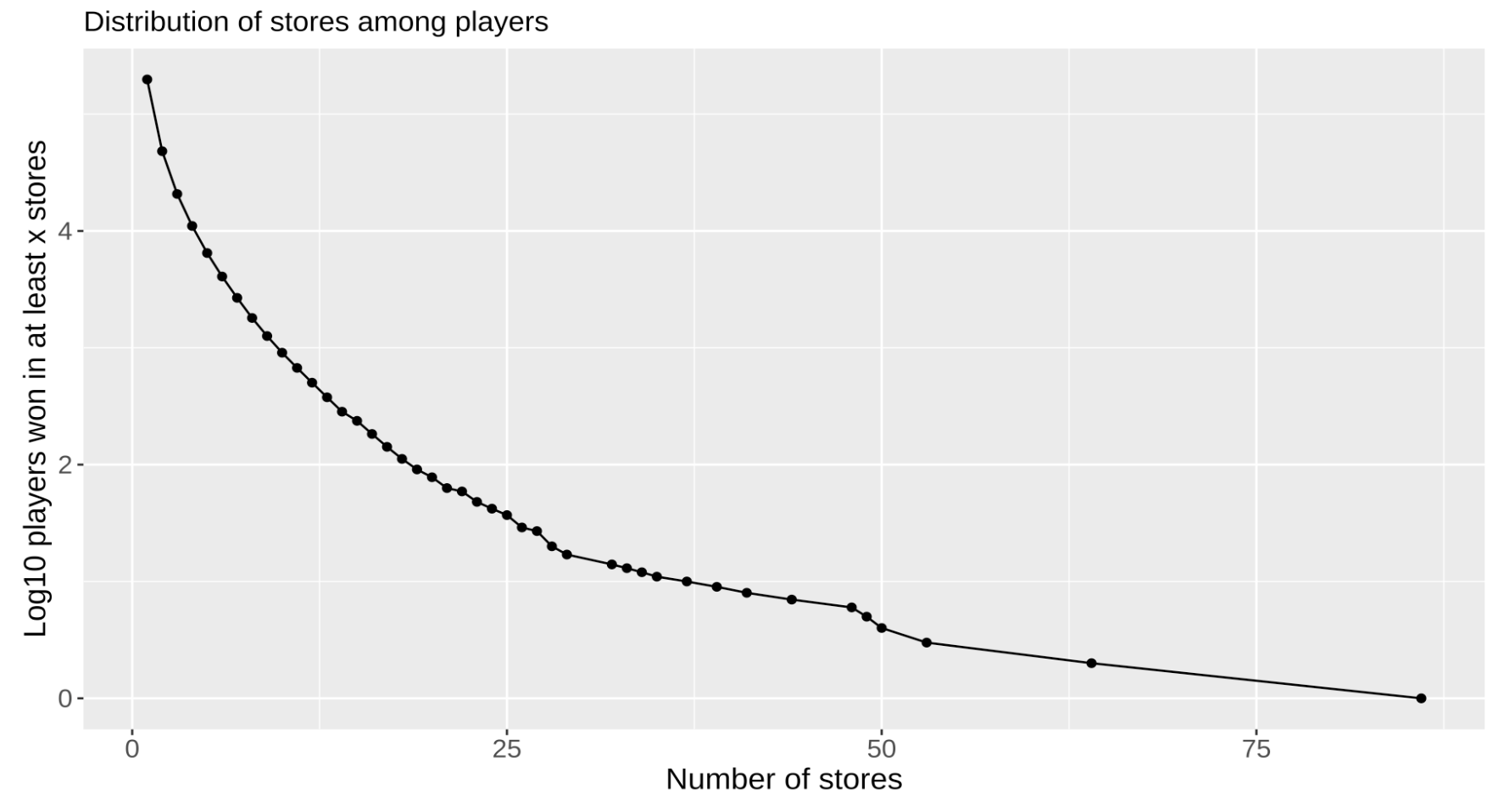}
\caption{Scatter plot of $\log_{10}$ of people who won in at least x stores (y axis) vs the number of stores (x axis). Each dot corresponds to the total number of winners who won a big prize at in least $x$ stores.}
\label{fig:DistributionStoreGraph}
\end{figure}
The distribution of stores is similar to the distribution of wins among players. Given that the majority of players experienced a single win, it is expected that they only won in a single store. As we move towards players with multiple wins, the number of individuals sharply decreases, with approximately 1,000 winners having obtained their big prizes from at least four different stores. Only around 100 players won big prizes in at least 18 different stores. Comparing to the data in Figure~\ref{fig:WinDisGraph} where around 100 players won around 49 times, we can conclude that many high-volume players are likely playing at only a few specific stores. In contrast, the most exceptional player in our dataset won prizes that were distributed across 86 different stores.

\section{Methods}
A naive approach to identifying unusual lottery activity would be looking for players with the most wins. However, different lotteries vary in chances to win large prizes. The Power Ball has a probability of winning over \$600 as low as $1.984933\times10^{-7}$, while the \$30 scratch-off -- Ultimate Millions has a chance to win at least \$600 over 0.0009. An Ultimate Millions player would win many more prizes above \$600 than a Power Ball player given similar frequencies of lottery purchasing. Since numbers of wins alone do not adequately measure behavior, we instead focus on net monetary winnings as our metric for how intensively people play the lottery. We predict the net winnings using a geometric distribution-based model that accounts for unrecorded small prizes (less than \$600). However, attempting to identify potentially suspicious players by estimated net winnings alone will still result in including both legitimate habitual players and discounters. 
Therefore, to identify persons who are suspected of ticket discounting, we also look into the players ticket buying behavior measured by entropy of the distribution of stores where their winning tickets were bought. See Figure ~\ref{fig:entroyGraph} for a plot of these two metrics for every player in the dataset. In order to assess whether individuals with high potential losses and high entropy are suspicious or simply exceptionally lucky, we investigate their winning pattern using a stochastic model.  \par


\subsection{
Estimation of mean net gain}
\label{sec:out-of-pocket gain}

When an individual participates in a lottery, they are purchasing tickets in the hopes of winning a large prize. The number of tickets they need to buy before achieving a big win is like a geometric distribution, with the probability $p$ of winning the large prize on a single ticket. However, we need to consider more then the number of tickets bought when considering average net winnings associated with a big prize. Individuals also win numerous unrecorded small prizes on the way to a big prize. Thus to calculate the overall net gain or loss for each lottery winner, we must consider the number of tickets purchased before winning a recorded prize and any smaller prizes (mless than \$600) the person may receive from those tickets. Since our goal is to first identify people who likely have outlying losses, we propose a simple and computationally efficient method to estimate expected values of those losses. A more realistic simulation based model is proposed in Section~\ref{sec:stochastic} to further investigate players identified by this simple method.
 
To account for small prizes, we find the overall return rate of lotteries in NC. The return rate ($R$) of a lottery is defined as the percentage of money that individuals anticipate gaining from a single lottery purchase. This rate is determined by dividing the total money won ($g_{all}$) from both big prizes (over \$600) ($g_{big}$) and small prizes (less than \$600) ($g_{small}$) by the total money spent on lottery tickets ($s_{all}$). However, since now we focus on calculating the losses incurred to win a single big prize, where the big prize amount is already known, we are mainly interested in the return rate of small prizes. Therefore, we define the small price return rate of a lottery ($R_s$) in our paper as the percentage of small prizes (less than \$600) that an individual can anticipate receiving from a single lottery purchase. This rate is calculated by dividing the total value of all prizes by the total amount of money spent on lottery tickets, while subtracting the sum of prizes exceeding \$600, 
 \begin{equation*}
R = \frac {g_{all}}  {s_{all}}, \qquad
g_{small} = g_{all} - g_{big}, \qquad
R_s = \frac {g_{small}}  {s_{all}}.
\end{equation*}

\begin{figure}[H]
\centering
\includegraphics[width=0.8\textwidth]{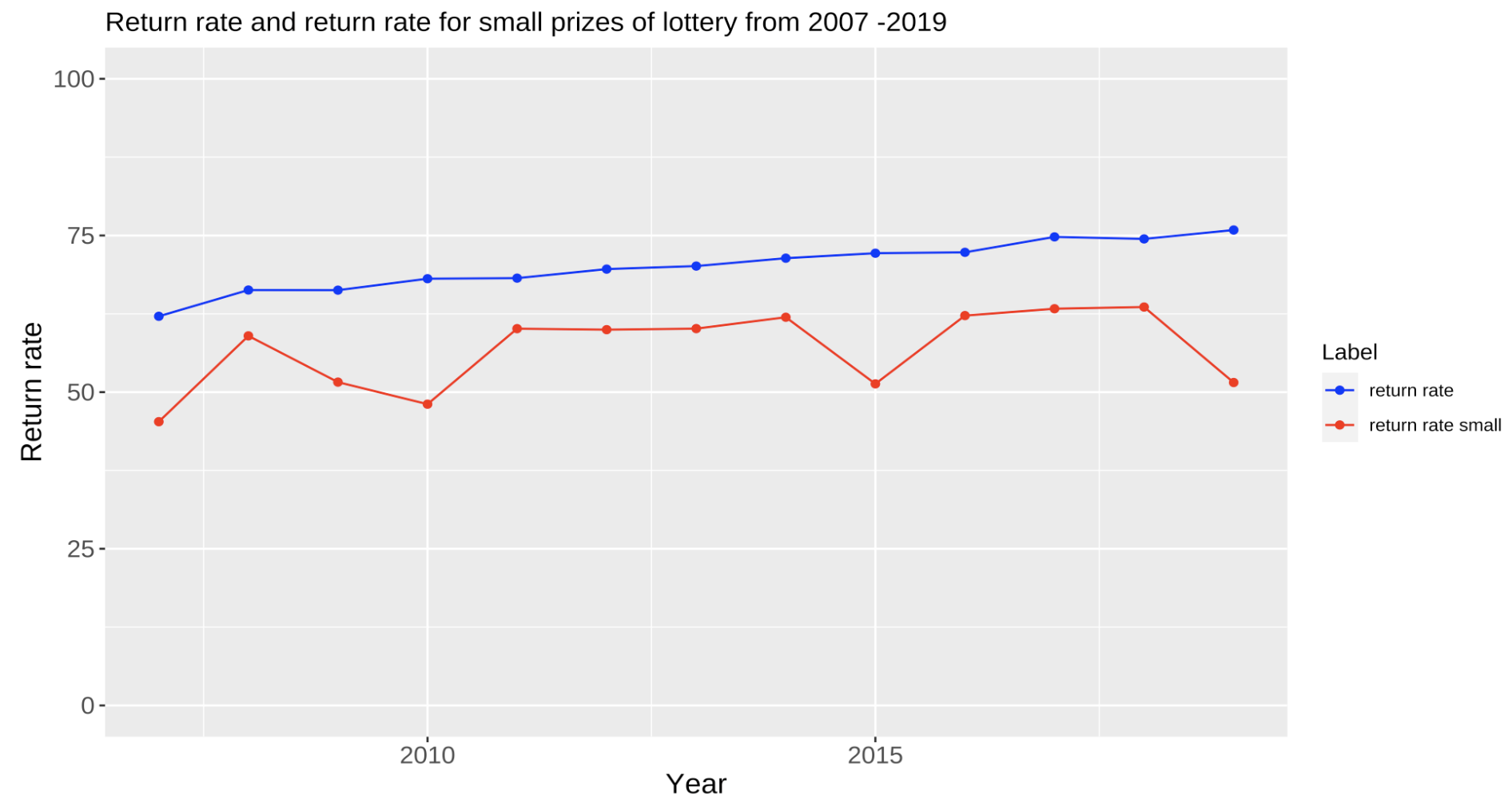}
\caption{Return rate and return rate for small prizes (in \%) of NC lottery from 2007 - 2019. For example, for every dollar spent on NC lottery tickets in 2007, about 70 cents were returned to customers in the form of prizes out of which about 55 cents were prizes less then \$600.}
\label{fig:ReturnRateGraph}
\end{figure}

 We investigated  the overall lottery return rate of NC and calculated the return rate of NC lotteries from 2007-2019 (the same full year time range of our dataset) according to the lottery report from the United States Census Bureau and the big prizes recorded in our dataset. Figure~\ref{fig:ReturnRateGraph} shows a graph of both the overall return rate ($R$, in blue) and the return rate for small prizes ($R_s$, in red) from 2007 to 2019. As can be seen from Figure~\ref{fig:ReturnRateGraph}, the return rate of unrecorded prizes for commonly played lottery games is nontrivial, meaning that the number of small prizes won could be significant before a winner wins a recorded prize.

  We use the mean of the geometric distribution to calculate the expected number of tickets needed ($E[N_{i,j}]$) to win one prize ($j$) for one player ($i$) of over \$600. We use it to model the number of tickets an individual is expected to purchase before achieving the win of over \$600. The probability ($p$) of winning over \$600 is calculated by averaging the probability of winning over \$600 from 42 different types of instant scratch-offs, with prices ranging from \$2-\$30. The specific probability used is 0.001226816. Similarly, the cost of a single lottery ticket ($\bar C$), is determined by averaging the costs of the 42 types of instant scratch-offs. The expected  return rate from small prizes $E[{R_s}]$ is estimated as $0.5677$ using the average return rate for small prizes from Figure~\ref{fig:ReturnRateGraph}. The recorded prize won on a certain record $j$ in the winning history of player $i$ is denoted as  $P^b_{i,j}$. Thus, the {\em mean net gain} ($E[G_{i,j}]$) of one single recorded win ($j$) of a particular player ($i$) is:
\begin{equation}
E[G_{i,j}] = P^b_{i,j} - E[N_{i,j}]*{\bar C}*(1-E[{R_s}])
\label{eq:geometric-model}
\end{equation}
We compute the estimated mean net gain for every recorded prize $j$ of each player. The total mean net gain ($E[G_{i}]$) for one player is the sum of the mean net gains for each recorded prize. For example, if a winner won a \$600 prize, the expected net gain for that prize would be $600 - (1/0.001226816) \times 14.32653 \times (1-0.5677) = -4448.319$.

The resulting mean net gains vary by several orders of magnitude among players in our data set. Therefore we employ a logarithmic transformation in graphical displays involving mean net gain, e.g. Figure~\ref{fig:entroyGraph}. In particular we will plot the {\em log mean net loss}
Notice, that players who are estimated to make money have their log loss displayed as 0 in Figures~\ref{fig:entroyGraph} and \ref{fig:entroyClusterGraph}.

\subsection{Entropy}
\label{sec:entropy}
As observed in Figure~\ref{fig:DistributionStoreGraph}, among players who have won more than once, the number of wins they have is considerably lower than the number of stores in which they have won big prizes. This suggests that a significant portion of the big players exhibit a preference for certain stores when purchasing lottery tickets, rather than choosing points of purchase in their vicinity at random. Therefore, players with many apparent wins across many stores are more likely to be potential ticket discounters. We quantify the range of lottery purchasing behaviors using entropy of the distribution of wins per store. Large entropy may be indicative of a suspicious player. The entropy (${E_i}$) for each player($i$) is defined as:
\begin{equation}
    {E_i} = -\sum_{n=1}^{N}{\left(\frac{W_{in}}{W_i}\right)}\log{\left(\frac{W_{in}}{W_i}\right)},
    \label{eq:entropy}
\end{equation}
where ${W_{in}}$ is the number of wins in a store $n$ for player $i$, $W_i$ is the total wins of the player $i$, and $N$ is the total number of distinct stores in which player $i$ won big prizes.

\begin{figure}[H]
\centering
\includegraphics[width=0.8\textwidth]{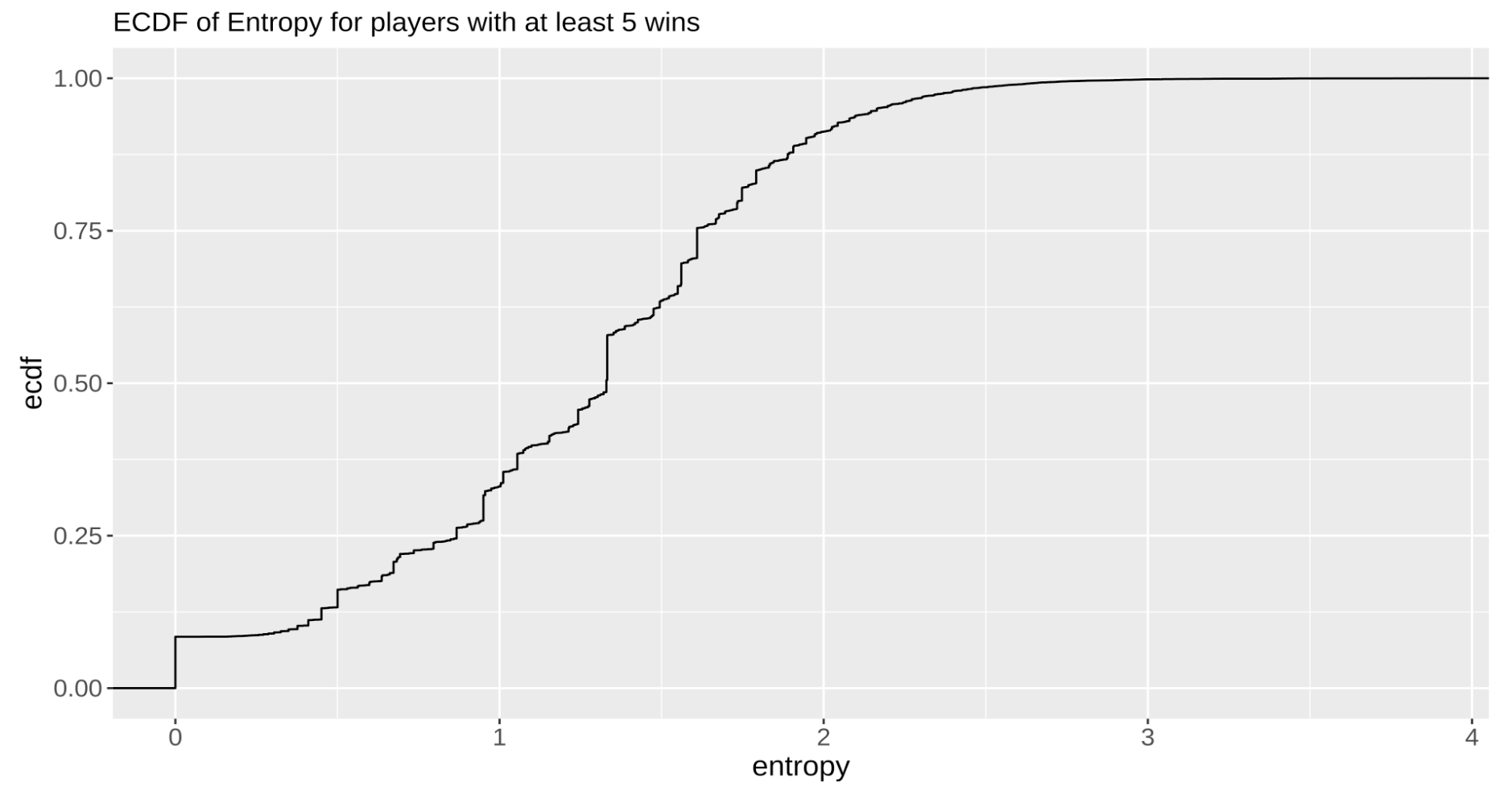}
\caption{Empirical Cumulative Distribution of Entropy for players with at least 5 wins. }
\label{fig:ECDFEntropy}
\end{figure}

Figure~\ref{fig:ECDFEntropy} shows the empirical distribution function (ECDF) of the entropy values $E_i$ for all players with at least 5 wins. Most entropy values are relatively small indicating concentrated distribution of the stores where players purchased their winning tickets. Recall that the uniform distribution on $N$ points has entropy of $\log(N)$. Moreover, close to 10\% of these frequent winners purchased all their winning tickets in one store.

\subsection{Stochastic model for net gain}
\label{sec:stochastic}

While the method in Section~\ref{sec:out-of-pocket gain} is an computationally efficient method for estimating net monetary gain from the lottery for each player, this method made a number of simplifying assumptions. Additionally, we also want to be able to estimate potential stochastic variation among the players deemed potentially suspicious by the net gain and entropy metrics. This would allow us to account for any inadvertent inclusion of unusually lucky individuals in the detection procedure. Therefore, we propose a stochastic model that simulates the actual experience of playing the lottery according to the probability of prizes for each lottery game. 

For the purposes of our analysis, we assume that the result of each instance of buying a lottery ticket can be treated as an independent event. This is clearly true for online lottery games such as Pick 4, Pick 3, Powerball, etc. because they are based solely on the numbers selected using the state's random number generator. If a player buys a scratch-off ticket, there will be one less ticket in the lottery ticket pool causing the probability to win the recorded prize to change. However, given the substantial number of tickets printed for any given scratch-off lottery (\url{https://nclottery.com/scratch-off}, accessed on 9/2/2023), we can reasonably overlook any negligible fluctuations in the probability of winning a prize from a scratch-off pool over time and assume independence for scratch-offs as well.  

For each player $i$, we define the recorded prize won on a certain winning ticket $j$ in their winning history to be $P^b_{i,k,j}$, and the net gain for winning that single recorded prize ($P^b_{i,k,j}$) to be $G_{i,k,j}$, where $k=1,\ldots K$ are the replicate runs of the simulation model. For each recorded ticket, we know the type of lottery played, the amount won $P^b_{i,k,j}$, the associated ticket cost $C_{i,j}$, and the probabilities of winning any prize, big or small. Thus we propose simulating all purchases leading up to each recorded prize, and tabulating all small prizes ($P^s_{i,k,j,x}$) won along the way. Here we are assuming that the player purchases tickets from the same lottery until they win a large prize. Once a purchase results in a prize greater than \$600, the simulation for this recorded prize halts. We capture the number of simulated tickets purchased $N_{i,k,j}$ and the total amount of simulated small prizes $ \sum_{x}P^s_{i,k,j,x}$. 
The simulated net gain associated with this ticket for one simulation run are given by
\begin{equation}
\begin{split}
G_{i,k,j} = P^b_{i,k,j} + \sum_{x}P^s_{i,k,j,x} - N_{i,k,j} * C_{i,j}
\end{split}
\label{eq:stochastic-model}
\end{equation}
If an individual won a single \$600 prize, we would continue drawing and record any additional smaller prizes they accumulate until one time the prize value exceeds \$600. For instance, let us consider a scenario where the person bought 100 \$10 lottery tickets before winning the \$600 prize. During these 100 lottery draws, they only received 2 \$20 prizes. The final total for the person's \$600 prize would be calculated as follows: $600 + 2 \times 20 - 10 \times 100 = -360$.

In contrast to \eqref{eq:geometric-model}, the number of tickets purchased and the total value of small prizes in \eqref{eq:stochastic-model} are generated using simulation rather than expected value. The total net gain for a player is calculated by summing over the recorded prizes $j$: 
\begin{equation*}
\begin{split}
G_{i} = \sum_{j}{G_{i,j}}.
\end{split}
\end{equation*}

For each player studied, the model is run 60,000 times. Because obtaining the full range of prize probabilities for each lottery is prohibitively complex, we select one representative lottery with approximately average win probability at each ticket price. The aggregated small prize probability and recorded prize probability from the representative lotteries are used for each simulated ticket purchased. The selected lotteries and the prize probability for the representative lotteries can be seen in the supplementary material. Since the model is run 60,000 times for a single player, we have 60,000 different estimated net gains for each player. We summarize the simulation results by reporting the net gain, as well as a 80\% simulation based confidence interval based on the 60,000 simulated net gains. As we focus on investigating players who exhibit exceptional success among all habitual players, it is crucial to consider a multiple testing adjustment when reporting the confidence interval for big players. In particular, we have applied the Bonferroni Adjustment to all the big players.

In order to use a Bonferroni adjustment, we need  to estimate the number ($B$) of big players present in the dataset. Specifically, we choose $B$ to be the number of individuals whose entropy (calculated in Section~\ref{sec:entropy}) exceeds the threshold of winning 5 big prizes in 5 distinct stores, i.e., the number of individual winners with entropy \eqref{eq:entropy} larger than $\log(5)$. This threshold selects every high-volume player with buying habits that span a large number of stores.
Using our dataset, the value of $B$ is determined to be 4320. Consequently, we will report the adjusted 10th percentile using the $10/4320=0.0023148$ percentile, while the adjusted 90th percentile will use the $100-10/4320=99.99769$ percentile of the 60,000 simulated net gains as the lower and upper bound of the 80\% simulation based confidence interval.

\section{Results}
\subsection{Initial screening results}

\begin{figure}[H]
\centering
\includegraphics[width=0.8\textwidth]{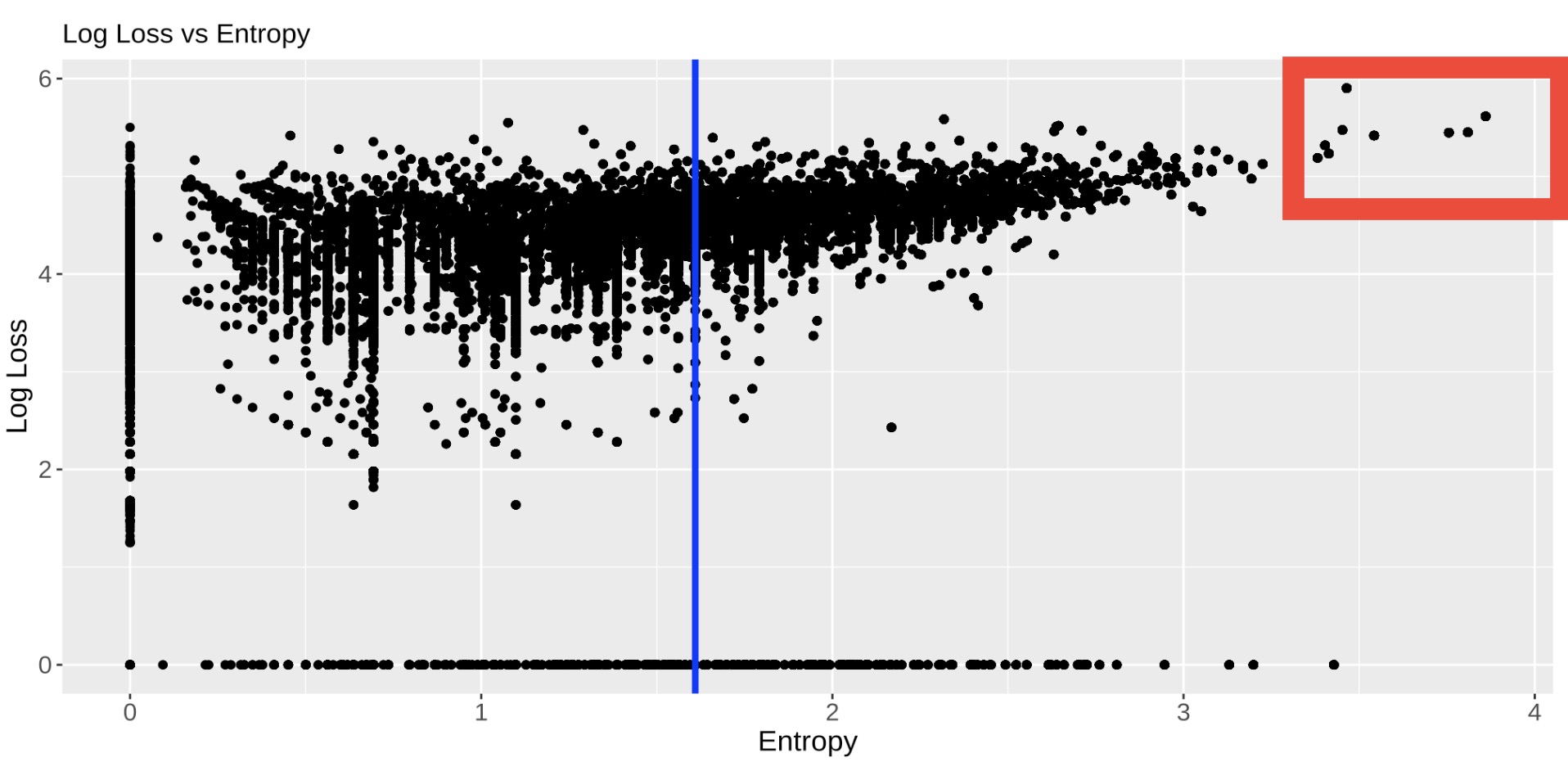}
\caption{Scatter plot of $\log_{10}$ of estimated net losses (with players who made money shown as 0) on NC education lottery (y axis) vs entropy of store distribution where winning tickets were bought(x axis). Each dot corresponds to an individual who won at least one prize of \$600. Zero $y$-value corresponding to people who were estimated to make money. The red box shows nine suspicious individuals with both large losses and high entropy. The blue line shows the entropy threshold we use for the Bonferroni Adjustment.}
\label{fig:entroyGraph}
\end{figure}

As discussed at the end of Section~\ref{sec:out-of-pocket gain}, the calculated mean net gains varied across multiple orders of magnitude, so we display the values using a base 10 logarithm transformation. We visually inspect the data for people with both large losses and suspicious store buying behavior by plotting log mean net loss and entropy in a scatter plot (Figure~\ref{fig:entroyGraph}). The greater the losses and entropy, the more suspicious the person appears. The correlation between the log loss and entropy for players with at least five wins is approximately 0.12, indicating a weak association between these two factors.
We identified nine outliers by taking the nine winners with the largest losses and entropy in the upper right corner (the red square in Figure~\ref{fig:entroyGraph}), indicating these players seem to lose a lot of money playing the lottery and go to many different stores to buy tickets. We flagged these nine suspicious winners for a further investigation.

\subsection{Stochastic model results}

We ran the simulation model described in Section~\ref{sec:stochastic} on the nine players we identified as unusual in Figure~\ref{fig:entroyGraph} to estimate the range of money they might have spent. For each single win, we simulated 60,000 instances and rounded the results to the nearest thousand.

\begin{table}
\centering 
\begin{tabular}{|p{1.9cm}|p{1.9cm}|p{1.9cm}|p{1.9cm}|p{1.9cm}|p{1.9cm}|p{1.9cm}|}
 \hline
  Name& Number of wins & Total reported winnings & Mean net gain & 10 percentile net gain& 90 percentile net gain\\
 \hline
 Winner 1 & 78 & \$114K & -\$715K & -\$1150K & -\$387K\\
 \hline
 Winner 2 & 76 & \$102K & -\$550K & -\$1010K & -\$274K\\
 \hline
 Winner 3 & 277 & \$601K & -\$1496K & -\$2084K & -\$948K\\
 \hline
 Winner 4 & 68 & \$82K & -\$482K & -\$854K & \$41K\\
 \hline
 Winner 5 & 154 & \$366K &-\$673K & -\$1078K & -\$344K\\
 \hline
 Winner 6 & 76 & \$86K &-\$668K & -\$1123K & -\$157K\\
 \hline
 Winner 7 & 58 & \$86K & -\$515K & -\$889K & -\$258K\\
 \hline
 Winner 8 & 45 & \$57K & -\$418K & -\$1275K & -\$3K\\
 \hline
 Winner 9 & 53 & \$113K & -\$245K & -\$557K & \$150K\\
 \hline
\end{tabular}
\caption{Net gain estimated using the Bonferroni adjusted stochastic model for the nine suspicious players indicated in the red box of Figure~\ref{fig:entroyGraph}. Prizes people won from the lottery are marked as positive. The money people lost in the lottery are marked as negative.}
\label{tab:Top Nine simulation costs}
\end{table}
As shown in Table~\ref{tab:Top Nine simulation costs}, each of these unusual players except Winner 4, 8 and 9 would have needed to spend several hundred thousand dollars even in the best-case scenario to win so many times in the lottery. Considering these people also have high entropy, we might conclude that they bought tickets from other people. In the cases of Winner 4, and Winner 9, despite their large potential losses in terms of mean and 10th percentile, their 90th percentiles does not exclude potential positive gains. The underlying explanations for why these players' simulation based prediction intervals are so wide will be discussed in Section~\ref{sec:discussion}.

\subsection{K-means clustering}
\label{sec:kmeansmethod}
Upon analyzing the outcomes from Section 4.1 and Section 4.2, we observed that the players we flagged were close and isolated in Figure~\ref{fig:entroyGraph}, implying a particular lottery purchasing pattern within a specific group of players. In this section we utilize $K$-Means clustering to further investigate whether additional individuals exhibit similar lottery buying behaviors as the flagged players. 

To this end, we define a 6-dimensional feature vector for each player based on their winning ticket purchasing pattern across stores. The first five features are the proportions of winning tickets purchased at each of that player's five most-visited stores, and the sixth feature is the proportion of winning tickets purchased by that player at any other stores. For example, if a person purchased tickets from ten different stores and won two times at each store, that person's 6-dimensional feature vector would be $(0.1, 0.1, 0.1, 0.1, 0.1, 0.5)$.

The $K$-means clustering algorithm is then applied on the 6-dimensional feature vectors described above using several different total number of clusters $K$. With the exception of Winner~3, all the other suspicious winners are clustered together and this finding holds over a wide range of total number of clusters $K$.

\begin{figure}[H]
\centering
\includegraphics[width=0.8\textwidth]{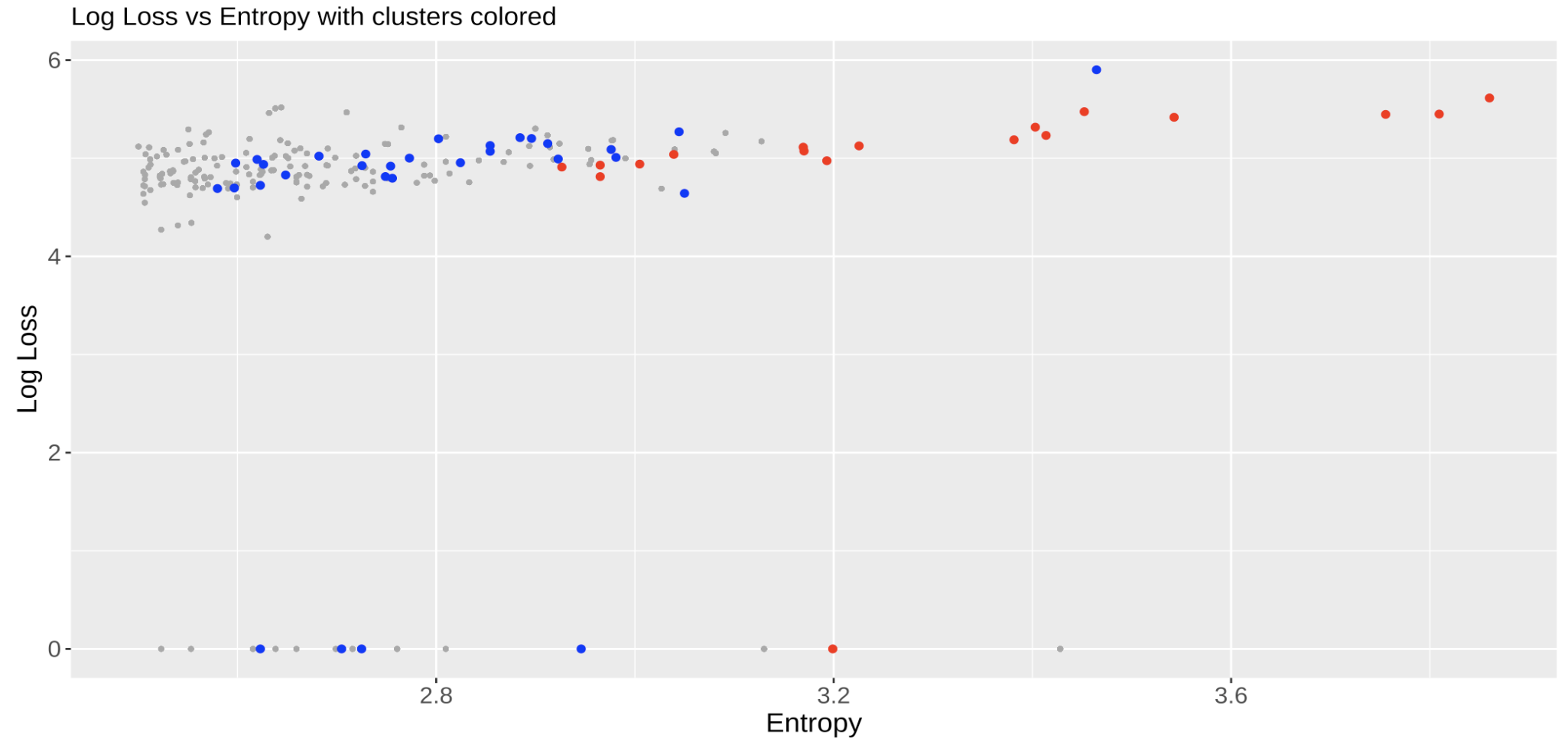}
\caption{Zoomed in scatter plot of $\log_{10}$ of estimated net losses (with positive gains shown as 0) on NC education lottery (y axis) vs entropy of store distribution where winning tickets were bought(x axis). Each dot corresponds to a individual who won at least one prize of \$600 or more and has an entropy bigger than or equal to 2.5. The red points on the graph indicate players who belong to the same cluster as the nine suspicious players except Winner~3. The blue points represent players who are in the same cluster as Winner~3.}
\label{fig:entroyClusterGraph}
\end{figure}

In Figure~\ref{fig:entroyClusterGraph} we present the clustering results computed using $K=25$.
Two clusters including the unusual players are marked with the red and blue plotting characters. Within the nine previously identified suspicious winners identified by the red square in Figure~\ref{fig:entroyGraph}, Winner 3 is contained within the blue cluster, whereas all the rest are located in the red cluster.
As can be seen in the graph, most people in the red cluster have high entropy and high losses, meaning they are all potentially suspicious. However, the red cluster also contains some potentially lucky players with positive mean net gains.

\begin{table}
\centering 
\begin{tabular}{|p{1.9cm}|p{1.9cm}|p{1.9cm}|p{1.9cm}|p{1.9cm}|p{1.9cm}|p{1.9cm}|}
 \hline
  Name& Number of wins & Total reported winnings & Mean net gain & 10 percentile net gain& 90 percentile net gain\\
 \hline
 Winner 11 & 40 & \$1168K &\$-9059K & \$-87564K & \$818K\\
 \hline
 Winner 12 & 27 & \$42K &\$-245K & \$-511K & \$-674K\\
 \hline
 Winner 13 & 24 & \$34K &\$-222K & \$-489K & \$-63K\\
 \hline
 Winner 14 & 22 & \$26K &\$-209K & \$-468K & \$-53K\\
 \hline
 Winner 15 & 34& \$38K &\$-317K & \$-610K & \$-111K\\
 \hline
 Winner 16 & 34 & \$42K &\$-314K & \$-629K & \$-114K\\
 \hline
 Winner 17 & 30 & \$33K &\$-282K & \$-548K & \$-112K\\
 \hline
 Winner 18 & 22 & \$46K &\$-126K & \$-316K & \$-14K\\
 \hline
 Winner 19 & 20 & \$20K &\$-202K & \$-475K & \$-54K\\
 \hline
 Winner 20 & 27 & \$29K &\$-275K & \$-535K & \$-99K\\
 \hline
\end{tabular}
\caption{Net gain estimated using the Bonferroni adjusted stochastic model for the additional suspicious players indicated as red dots outside the red box in Figure~\ref{fig:entroyGraph}. The columns are the same as in Table~\ref{tab:Top Nine simulation costs}}
\label{tab:Red cluster winners with top nine suspicious winners estimate costs}
\end{table}
To investigate the red cluster further, we repeated the simulation for all the remaining 11 people that were not included in the original 9 players studied in Table~\ref{tab:Top Nine simulation costs}. The additional simulation results are provided in Table~\ref{tab:Red cluster winners with top nine suspicious winners estimate costs}. 
Because with exception of Winner~11 the upper bounds of the simulation-based prediction intervals are negative, we can be highly confident that these winners have lost large sums of money if they indeed purchased their tickets from the NCEL. Since these players also exhibit an unusual pattern of stores where winning tickets were purchased we have a strong suspicion that these people bought winning tickets from other people.

\section{Discussion}
\label{sec:discussion}
As is shown in the stochastic model results, the majority of players in the suspicious cluster displayed substantial losses even in the best-case scenario if their wins came from legitimate ticket purchases. Combined with their high entropy, this leaves them looking suspicious as potential ticket discounters. However, it is worth noting that among the suspicious players, there are three winners who exhibit potential positive gains at the top end of the range of simulated outcomes from the model. The reason for that lies in their extensive participation in online lottery games such as Pick 4 and Powerball. These online games have a low probability of winning prizes exceeding \$600, while maintaining relatively low ticket prices, resulting in comparatively unpredictable outcomes relative to players that play other lottery games. In some instances, players may have experienced extraordinary luck, winning a significant amount while only spending a minimal sum on tickets. Consequently, this wide range of outcomes produces relatively wider uncertainty intervals for these players. Despite their potentially positive net gain as evidenced by the Bonferroni adjusted 90th percentile, it is important to consider that all of these players still have remarkably high entropy values and display substantial losses on average. Therefore, one may still choose to consider Winners 4, 8, 9, and 11 as potentially suspicious players.

In conclusion, we associated estimated net gain with store buying behaviors to investigate suspicious lottery players. Through our initial analysis that utilized the geometric distribution, stochastic models, and entropy, we identified nine suspicious winners with both large losses and high entropy. Using cluster analysis, we were able to identify fourteen additional suspicious winners who shared similar purchasing habits to the initial nine. As we did not consider geographic location in our algorithm, future work may incorporate geographic location in the analysis of store buying behavior. Also, a new analysis could be performed with a focus on stores where many winning scratch-off tickets were purchased with the aim of identifying potential fraud by store owners and clerks.

\section*{Supplementary Material}

\begin{description}

\item[Code for all the implementation of method:] R Markdown file containing code to perform all the method discussed in the article. (.rmd file)

\item[North Carolina Education Lottery public record:] Dataset described in Section~2. (.RData file)

\item[Representative lotteries with price probabilities for stochastic model:] A table containing the representative lottery names, prizes, and probabilities for each prizes used in the stochastic model in Section~4.2. (.xlsx file)

\item[Pick 4 with prize probabilities for stochastic model:] A table containing the prizes and probabilities for each prizes of Pick 4 used in the stochastic model in Section~4.2. (.xlsx file)

\item[Detailed lottery information about the top nine outliers:] A zip file containing nine Excel files about detailed lottery information played by the nine outliers marked in Section~4.1.\footnote{Created by searching each lottery information from~\url{https://nclottery.com.} Unzip the file before use.} (.zip file)

\end{description}

\bibliography{Bibliography-MM-MC, bibtex.bib}
\bibliographystyle{apalike}

\end{document}